# Control of Coercivities in (Ga,Mn)As Thin Films by Small Concentrations of MnAs Nanoclusters

K. Y. Wang[*]

*School of Physics and Astronomy, University of Nottingham, Nottingham NG7 2RD*

M. Sawicki

*Institute of Physics, Polish Academy of Sciences, al. Lotników 32/46, PL-02668 Warszawa, Poland*

K. W. Edmonds, R. P. Campion, A. W. Rushforth, A. A. Freeman, C. T. Foxon, and B. L. Gallagher

*School of Physics and Astronomy, University of Nottingham, Nottingham NG7 2RD*

T. Dietl[1,2,3]

[1]*Institute of Physics, Polish Academy of Sciences, al. Lotników 32/46, PL-02668 Warszawa, Poland*

[2]*ERATO Semiconductor Spintronics Project Japan Science and Technology Corporation 02668 Warszawa, Poland*

[3]*Institute of Theoretical Physics Warsaw University 00681 Warszawa, Poland*

**Abstract**

*We demonstrate that low concentrations of a secondary magnetic phase in (Ga,Mn)As thin films can enhance the coercivity by factors up to ~100 without significantly degrading the Curie temperature or saturation magnetisation. Magnetic measurements indicate that the secondary phase consists of MnAs nanoclusters, of average size ~7nm. This approach to controlling the coercivity while maintaining high Curie temperature, may be important for realizing ferromagnetic semiconductor based devices.*

The (III,Mn)V ferromagnetic semiconductor family [1] has attracted much attention for its potential applications in spintronics [2], in which, for examples, logic and memory operations may be integrated on a single device. (Ga,Mn)As, with Curie temperature, $T_C$, as high as 173 K [3], is one of the most promising candidates for such applications. A key parameter for any spintronics device is the coercivity $H_C$ of the magnetic layers. In (Ga,Mn)As, $H_C$ is typically found to vary between 120 Oe and 5 Oe (T ≥ 5 K), generally decreasing with increasing hole density and increasing $T_C$ [4]. Such soft magnetic material makes the manipulation of magnetisation easy but it is not particularly suitable for applications requiring non-volatility. The coercive fields required for magnetic storage media are typically between 350 Oe to 4000 Oe. While exchange biasing of (Ga,Mn)As using antiferromagnetic layers has recently been demonstrated [5], $H_C$ typically remains below 200 Oe, and poor reproducibility is reported [5]. In order to realize non-volatility, spin information transmission and manipulation in the same material system, it is important to develop methods for control of coercivities of ferromagnetic semiconductors such as (Ga,Mn)As.

High quality (Ga,Mn)As must be obtained by non-equilibrium low temperature molecular-beam epitaxy (LT-MBE) [1,6]. Growth or post-growth annealing at higher temperatures [7,8] leads to phase separation into MnAs clusters embedded in the relaxed GaAs host matrix. The growth temperature must therefore be carefully controlled to prevent phase segregation while also minimising As anti-site formation. The onset of phase segregation can be identified by a transition from 2D to 3D RHEED patterns during growth [9,10]. Here, we show that remarkably large coercivities can be obtained in (Ga,Mn)As thin films in which very low, but still detectable, levels of magnetic secondary phases are present.

---

[*] Now at Hitachi Cambridge Laboratory, Cambridge CB3 0HE, UK, E-mail:kwang@phy.cam.ac.uk

A wide range of 50 nm thick $Ga_{1-x}Mn_xAs$ films with $0.011 \leq x \leq 0.09$ were grown on semi-insulating GaAs(001) substrates by LT-MBE using $As_2$, where the layer structure is 50nm $Ga_{1-x}Mn_xAs$ / 50nm LT-GaAs / 100nm GaAs / GaAs(001). The growth temperature $T_g$ of the (Ga,Mn)As films and the LT-GaAs buffer was kept close to phase boundary between 2D and 3D growth. $T_g$ varied between around 200ºC-300ºC, decreasing with increasing Mn flux. Details of the growth are given elsewhere [10]. Mn concentration was determined from the Mn/Ga flux ratio, calibrated by secondary ion mass spectrometry (SIMS) measurements on 1μm thick films. Some of the samples were annealed in air at 190ºC for 50-150 hours, which is an established procedure for increasing the $T_C$ of thin (Ga,Mn)As films [11]. The magnetic properties of the layers were measured using a SQUID magnetometer.

Magnetization versus external magnetic field at 5 K for two (Ga,Mn)As thin films (sample A and sample B) with $x = 0.078$ is shown in Fig.1 (a) and (b), respectively. These samples were grown under similar growth conditions close to the 2D-3D phase boundary. After annealing sample A has a Curie temperature of 150K while that of sample B is 133K. Measurements along the orthogonal in-plane [110] and [1$\bar{1}$0] directions are shown, both before and after low temperature annealing. The samples show the magnetic uniaxial anisotropy between the [110] and [1$\bar{1}$0] directions, as is universally observed in (Ga,Mn)As thin films [12]. The most remarkable difference between these two samples is the coercivity, which is much larger for sample B than for sample A. Furthermore, $H_C$ decreases on low-temperature annealing of sample A, but increases for sample B. After annealing, $H_C$ is ~100 times larger for sample B than for sample A.

Magnetization curves recorded at 250K, well above the $T_C$ of the (Ga,Mn)As layers, are shown in Fig. 2. The M(H) curve for sample A shows (nearly) no field dependence. However, for sample B a clear ferromagnetic-like signal is observed, suggesting the presence of a secondary magnetic phase in this sample. This implication is further confirmed by the magnetization versus temperature under an external field of 1000 Oe for both samples, which is shown in the inset of Fig. 2. For sample A, the magnetic signal falls to zero a few K above its $T_C$. However, a non-zero magnetic signal persists till about 300 K for sample B. This temperature is close to already reported $T_C$ of MnAs grown on GaAs (001) (313 K) [13] or to MnAs precipitates in GaAs ($T_C \sim$ 350 K) [8], but is much lower than the $T_C$ of GaMn embedded in GaAs which is over 400K [14]. We conclude that the secondary magnetic phase is most likely to be composed of MnAs precipitates. Removing the top 15nm of sample B by etching in $H_2SO_4:H_2O_2:H_2O$ was found to have little effect on $T_C$, the magnetic anisotropy, or the ratio of the magnetic signals. Therefore, it appears that the MnAs precipitates are uniformly dispersed through the (Ga,Mn)As film.

Figure 3 shows the coercivity measured at 5 K versus the Curie temperature for the series of (Ga,Mn)As films. For the single-phase samples, $H_C$ varies from around 5 Oe to 120 Oe, and generally decreases with increasing $T_C$, similarly to previous studies [4]. The samples in which a secondary magnetic phase is observed show greatly increased $H_C$, which is found to increase nearly linearly with $T_C$. We find $T_C$ after annealing to be only around 10% smaller in the two-phase samples than in single phase samples with the same nominal Mn concentration, so that the very large values of $H_C$ are achieved with only a small detriment to $T_C$. The conductivity is also slightly lower in the two-phase samples. The room temperature conductivities before and after annealing are respectively 210 $\Omega^{-1}cm^{-1}$ and 520 $\Omega^{-1}cm^{-1}$ for sample A, and 180 $\Omega^{-1}cm^{-1}$ and 250 $\Omega^{-1}cm^{-1}$ for sample B.

No remnant magnetization is observed above the Curie temperature of (Ga,Mn)As in any of the samples. This indicates that the MnAs precipitates are superparamagnetic with a blocking temperature $T_B$ less than or equal to the $T_C$ of the (Ga,Mn)As. Magnetization curves above the $T_C$ of the (Ga,Mn)As layer for sample B are shown in Fig. 4. These can be described by

$$M = M_0(\coth[\frac{\mu H}{k_B T}] - \frac{K_B T}{\mu H}) + \frac{CH}{T - T_C} \qquad (1)$$

where the first term is the Langevin function for an assembly of non-interacting superparamagnetic particles above $T_B$ [15], and the second term accounts for the Curie-Weiss behavior of the paramagnetic (Ga,Mn)As film. $M_0 = n\mu/V_{total}$, n the number of MnAs particles and $V_{total}$ is the total volume of the (Ga,Mn)As thin film. The effective moment µ per MnAs particle is given by $M_S(T)<V_{MnAs}>$, where $M_S(T)$ is the saturation magnetization of MnAs and $<V_{MnAs}>$ the average particle volume. As shown in Fig. 4, the experimental results at different temperatures can be accurately reproduced by equation (1), with µ and n being temperature-dependent and temperature-independent fit parameters, respectively.

The obtained µ decreases with increasing temperature, as is shown in the inset of Fig. 4. Using the Brillouin function to fit this temperature dependence of the saturation cluster magnetization yields a Curie temperature of 310 ± 15 K, which is consistent with the findings presented in the inset of Fig. 2. At 150K the saturation magnetization is 2.9 ± 0.3 $\mu_B$ per Mn atom and the effective moment per super-paramagnetic cluster is 25000$\mu_B$, which gives an average MnAs particle volume of 270±50 nm$^3$, corresponding to a lateral size of ~6.5 ± 0.5nm. The concentration of

MnAs clusters is about $2.5\pm0.4 \times 10^{16}$ cm$^{-3}$, which comprise only $0.8 \pm 0.2\%$ of the volume of the epilayer.

The experimentally observed low coercivities of samples without secondary phase inclusions are much lower that those predicted for coherent rotation on the basis of the measured anisotropy constants [16] since the coercivity is controlled by domain wall nucleation and propagation [17]. We postulate that the MnAs nanoclusters act as pinning centers for the domain walls, strongly inhibiting their motion and thus reducing the efficiency of this reversal mechanism. Recent scanning Hall measurements seem to support this conjecture [18].

In summary, we have observed substantial enhancements of the coercivity of (Ga,Mn)As thin films containing secondary magnetic phases. This phase consists of nanometer size MnAs precipitates, which are uniformly distributed through the film and are superparamagnetic above the Curie temperature of the host (Ga,Mn)As. Such a large increase of $H_C$ by low levels of MnAs precipitates could be important for engineering (Ga,Mn)As-based multifunctional devices.

This research has been supported by EPSRC (GR/S81407/01), EU FENIKS(G5RD-CT-2001-0535), and Polish PBZ/KBN/044/PO3/2001.

**Figure Captions:**

Figure 1: The M(H) loops along [110](circles) and [1$\bar{1}$0](triangles) directions at 5K for the as-grown(solid) and annealed(open) $Ga_{0.922}Mn_{0.078}As$ (a) sample A and (b) sample . Note the change of horizontal scale between (a) and (b)

Figure 2 The M(H) curves at 250 K for both $Ga_{0.922}Mn_{0.078}As$ samples A and B; Inset: Temperature dependence of the magnetization measured along [110] direction for both annealed samples. The diamagnetic background has been removed in both graphs.

Figure 3: The measured coercive field at 5 K versus the Curie temperature for the studied as-grown and annealed (Ga,Mn)As single-phase samples (circles) and samples with second phase (squares).

Figure 4: The experimental (open symbols) and fitting (lines) M(H) curves at 150K (squares), 200K (circles) and 250 K (triangles) for annealed $Ga_{0.922}Mn_{0.078}As$ sample B; inset: the temperature dependence of the magnetic moment per MnAs cluster (Square symbols) obtained from equation 1, which is fitted with Brillouin function(line).

*Fig.1 K. Y. Wang et al.*

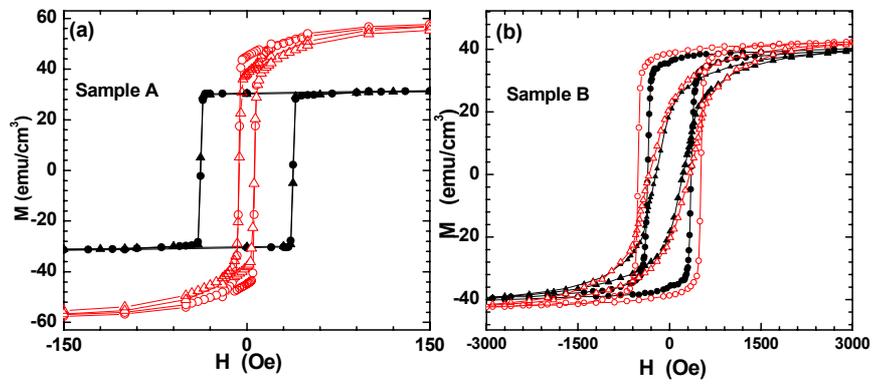

*Fig.2 K. Y. Wang et al.*

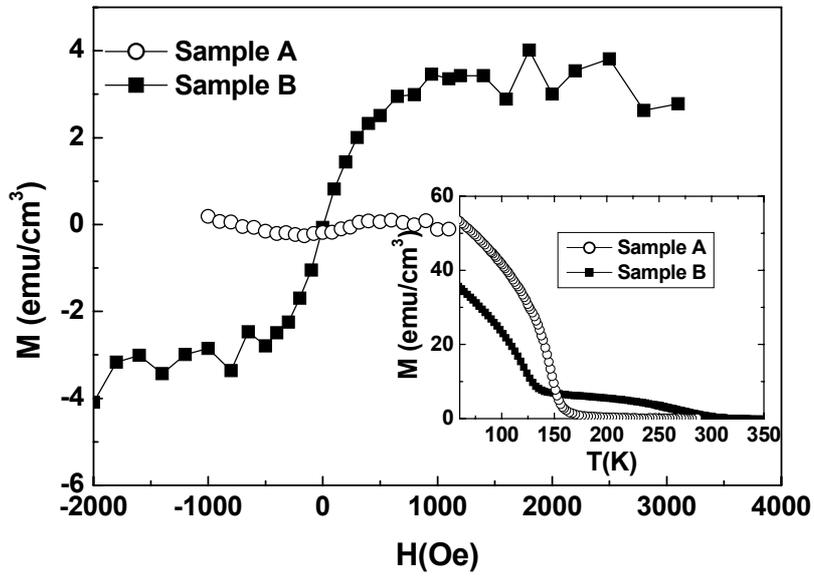

*Fig.3 K. Y. Wang et al.*

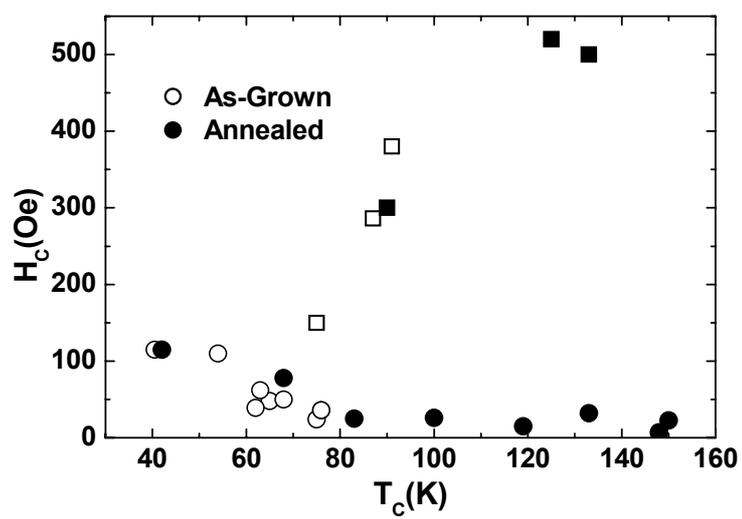

*Fig.4 K. Y. Wang et al.*

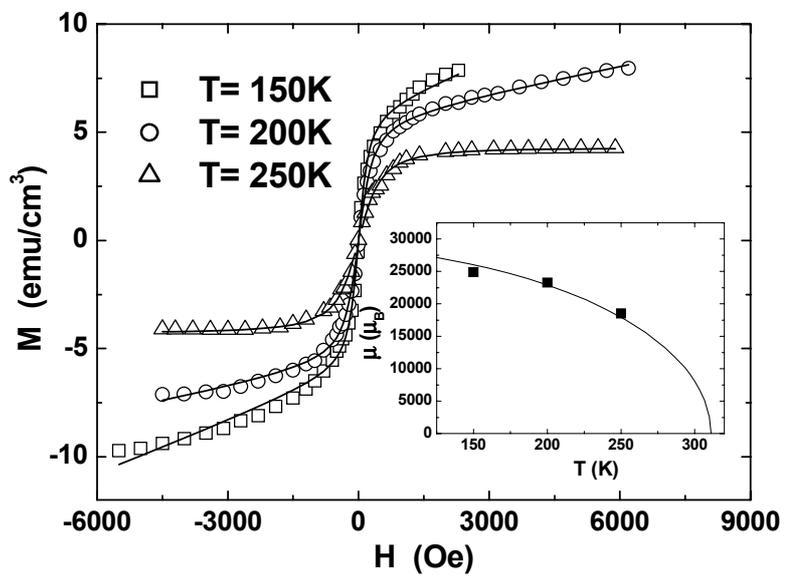